\documentclass[preprint,prX]{revtex4}
\usepackage{graphicx}
\usepackage{dcolumn}
\usepackage{bm}
\begin{document}
\preprint{APS}
\title{Universal Probability Distribution Function for Bursty Transport in Plasma Turbulence}
\author{{I.~Sandberg}$^{1,2}$, S.~Benkadda$^{3}$, X.~Garbet$^4$, G.~Ropokis$^2$,
K.~Hizanidis$^1$ and D. del-Castillo-Negrete$^5$}
\affiliation{$^1$ National Technical University of Athens, Association Euratom - Hellenic Republic\\
$^2$ National Observatory of Athens, Institute for Space Applications and Remote Sensing, Greece\\
$^3$ France-Japan Magnetic Fusion Laboratory, LIA 336/UMR 6633 CNRS-Universit\'e de Provence, Marseille, France\\
$^4$ I.R.F.M., CEA Cadarache, France\\
$^5$ Oak Ridge National Laboratory, Oak Ridge, Tennessee 37831-8071,
USA}
\date{\today}
\begin{abstract}
Bursty transport phenomena associated with convective motion present
universal statistical characteristics among different physical
systems. In this letter, a stochastic univariate model and the
associated probability distribution function for the description of
bursty transport in plasma turbulence is presented. The proposed
stochastic process recovers the universal distribution of density
fluctuations observed in plasma edge of several magnetic confinement
devices and the remarkable scaling between their skewness $S$ and
kurtosis $K$. Similar statistical characteristics of variabilities
have been also observed in other physical systems that are
characterized by convection such as the X-ray fluctuations emitted
by the Cygnus X-1 accretion disc plasmas and the sea surface
temperature fluctuations.
\end{abstract}

\maketitle

Plasma turbulence and the associated heat and particle transport
play a major role in the levels of plasma confinement in magnetic
fusion devices. In plasma edge the turbulent fluctuations are large
and bursty due to the convective motion of strongly nonlinear
structures formed during the nonlinear saturation of plasma
instabilities. Such coherent structures in an unambiguous manner
effectively contribute to radial transport and to intermittency
\cite{benkadda}. As a result, the transport process departs from the
diffusive picture associated with the Gaussian case of weak
independent fluctuations.

In order to understand the underlying mechanism of the turbulent
transport in the plasma edge, it is crucial to investigate the
statistical characteristics of these bursty fluctuations.
Experimental investigations have indeed revealed the bursty nature
of particle transport in the scrape--of--layer (SOL) of magnetically
confined plasmas \cite{Jha}. The appearance of structures such as
plasma blobs, avaloids \cite{AntarPRL} is attributed to the
formation of field-aligned structures - induced by the charge
separation of the magnetic curvature drifts - that propagate
radially far into the SOL.

\begin{figure}
\begin{center}
\includegraphics[width=80mm]{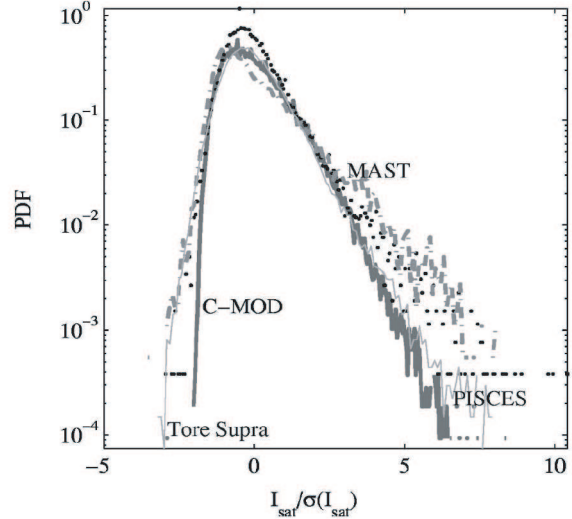} \vspace{-0.5cm}
\caption{{\small{ The PDF plot of the ion saturation current in the
Tore Supra (solid line), Alcator C-Mod (thick solid line), MAST
(dashed-dotted line), and PISCES (dots). The ion saturation current
was normalized to the standard deviation and the integral of the
four PDFs is set equal to 1. Figure reprinted with permission from
Ref. \cite{AntarPoP} (http://link.aip.org/link/?PHPAEN/10/419/1),
Fig. 3, Copyright 2003, American Institute of Physics. }}}
\label{fig:exppdf}
\end{center}
\end{figure}

The statistical behavior of the density fluctuations associated with
such bursty dynamics has been investigated in several experiments. A
comprehensive study \cite{AntarPoP}, that included measurements from
various magnetic confinement devices -- including TORE SUPRA, MAST
and ALCATOR C-MOD -- showed that the associated extreme probability
distribution functions (PDFs) are universal (see
Fig.~\ref{fig:exppdf}) in the sense that have the same properties in
many confinement devices with different configurations. The
investigations were carried out using Langmuir probes that measure
the saturation current.

Remarkably, similar \cite{dendy} extreme distributions have been
also observed for the bursty X-ray fluctuations associated with
transport events in the Cygnus X-1 accretion disc plasmas that are
linked to instabilities which give rise to turbulent transport and
extreme statistics (see Ref. \cite{greenhogh} and Fig. 12 therein).
However, known forms of extreme PDFs -- such as the Fr\'echet or the
Gumbel distributions -- do not have \cite{greenhogh} the proper form
to fit well the experimentally observed distributions associated
with bursty convective transport processes.

Bursty dynamics is characterized by strongly non-Gaussian PDFs and
non-vanishing probabilities of extreme events. In such dynamical
systems, the higher order moments are commonly used to determine the
scaling properties of the fluctuating fields. The non--Gaussian
features are usually quantified in terms of skewness and kurtosis of
the PDF of fluctuating fields. For a centered random variable
$\tilde{x}$, i.e. $\left<\tilde{x}\right>$=0, with  variance
$\sigma^2$, skewness is defined by the $S\equiv
\left<\tilde{x}^3\right>/\sigma^3$ and the kurtosis (in the newer
literature) by $K\equiv\left<\tilde{x}^4\right>/\sigma^4-3$.
Skewness is a measure of asymmetry of a PDF; if the left tail is
more pronounced than the right tail, the PDF has negative skewness
and when the reverse is true, it has positive skewness. Kurtosis
measures the excess probability (flatness) in the tails, where
excess is defined in relation to a Gaussian distribution. For
Gaussian distributions both $S$ and $K$ are equal to zero.

Using ten thousand observed density fluctuation signals measured in
TORPEX, Labit et al. \cite{Labit} showed that a unique parabolic
scaling relation holds, $K = 1.502 S^2-0.226$, between the skewness
$S$ and the kurtosis $K$ (see Figure 1 of Reference \cite{Labit}).
It was also shown that the PDFs of the measured signals, including
those characterized by a negative skewness can be described by a
special case of the Beta distribution. The density fluctuations were
associated with regimes of drift--interchange (D-I) turbulence
generated in regions of bad magnetic field curvature and convected
away by the $E \times B$ fluid motion.

Remarkably, there is a striking similarity of the observed $K$--$S$
scaling with that of sea surface temperature (SST) fluctuations that
are governed by advection through ocean currents \cite{Sura}. Sura
and Sardeshmukh \cite{Sura} proposed a nonlinear Langevin model that
can predict the observed scaling in some limits of the model
parameters while Krommes \cite{Krommes} generalized it to include
linear waves - an essential feature of D-I turbulence - and he
numerically calculated the associated PDF that includes four
indepedent parameters. The fact that the same scaling applies to
different physical situations leads to conjecturing that the scaling
arises due to basic constraints.
The general mathematical constraints on the $K-S$ relation, as arise
from the definition of kurtosis and skewness \cite{sattinPS}, do not
provide any insight about the observed scaling. A recent
comprehensive study on the observed $K-S$ scaling between various
fusion devices showed that the data align along parabolic curves
\cite{sattin}. A phenomenological model using the assumption that
the fluctuating signals include a linear combination of two basis
(Gamma or Beta ditributions) PDFs attempting to accommodate
experimental evidence was also provided. However, the large number
of the free parameters entering to the model and the small
difference between using one or a sum of two Beta PDFs make the
model less flexible \cite{sattin}.

It becomes evident that these observations among different physical
systems reveal a universal character associated with strongly
non--Gaussian processes. However, the key question is what kind of
underlying mechanism is responsible for the universally observed
statistical features and what PDF can describe them. In this Letter,
a univariate model for the statistical description of bursty
fluctuations is presented, using as an example the aforementioned
plasma edge density fluctuations. Furthermore, the associated PDF is
derived and it is shown that it recovers the universally observed
distributions, documented in Ref. \cite{AntarPoP} and the remarkable
parabolic scaling between skewness and kurtosis as well, documented
in Ref. \cite{Labit}. The derived results have universal character,
and thus may be applicable to all aforementioned physical systems.

\begin{figure}
\begin{center}
\includegraphics[width=80mm]{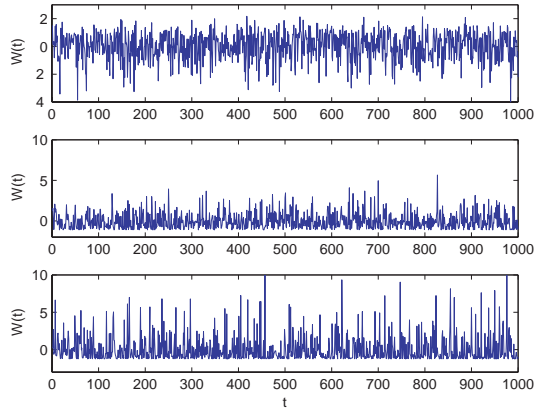}
\caption{{{(a) Example of bursty time--series $W(t)$ for different
values of $\gamma$: (a) -0.11, (b) 0.031, (c) 0.98, corresponding to
negative (top), positive (middle), and significantly large positive
(bottom) skewed distributions respectively. }}}
\label{fig:timeseries}
\end{center}
\end{figure}

The non--linear processes described by the standard models of
turbulence in magnetized plasma are quadratic and are linked with
small or large scale convection processes attributed to electric
drifts. Thus, it is natural one to assume that the universally
observed statistical characteristics associated with the bursty
behavior of fluctuations may be attributed to processes that emerge
from the non-linear quadratic interaction between turbulent fields.
However, extreme statistical features are expected to appear when
strong non--Gaussian processes coexist with Gaussian ones. In order
to describe the associated universal statistical properties, we
propose a univariate non-Gaussian process $W(t)$ given by:
\begin{equation} \label{ngproc_z1} W(t)=\frac{Z(t)-\left<Z(t)\right>}{\sigma_0}
+\gamma\frac{Z^2(t)-\left<Z^2(t)\right>}{\sigma_0^2} \end{equation}
which results from the superposition of a Gaussian $Z(t)$ (with
standard deviation $\sigma_0$) with a non-Gaussian process
attributed to the square of $Z(t)$. The latter corresponds to the
strongest quadratic non-linearity that may arise due to the
interaction of fluctuating fields. The coefficient $\gamma$, in
front of the quadratic non--Gaussian component is a parameter that
measures the deviation of $W(t)$ from Gaussianity. The process
$W(t)$ may well describe extreme bursty behavior of fluctuations
that is characterized by non-linear structures (blobs, avaloids)
that travel (convect) through a sea of Gaussian fluctuations. Figure
\ref{fig:timeseries} presents some typical time series $W(t)$
calculated from Eq. (1) by using a centered random Gaussian
fluctuation $Z(t)$. In all cases the bursty nature of $W(t)$ is
evident. For the sake of simplicity, from now on, we drop the
dependence of $Z$ and $W$ on time and consider that $Z$ is a
centered Gaussian process. 
It should be noted here that the process described by Eq. (1) has
been presented in the literature \cite{Lenschow} as a characteristic
simple-to-construct non-Gaussian process and was used as an example
for the calculation of the systematic errors of covariances and
moments up to fourth order of non-Gaussian time series. 

For the derivation of the PDF $p(w)$ of the random process $W(t)$,
we express the cumulative distribution function (CDF) $P(w)$ as
follows
\begin{equation}
\label{cdf} P(w)=\Pr \left( {W \le w } \right) = \Pr \left( {\gamma
\left( {Z - z_1 } \right)\left( {Z - z_2 } \right) \le 0} \right)
\end{equation}
where $z_1$ and $z_2$ are the roots of the polynomial
\begin{equation}
f(Z) = \gamma\frac{Z^2-{\sigma_0}^2}{{\sigma_0}^2} +
\frac{Z}{\sigma_0} - w.
\end{equation}
Equation \ref{cdf} can be expressed in terms of the CDF $P(Z)$ at
$z=z_1,~z_2$ and $p(w)$ can be derived by a simple differentiation
that leads to the following expression:
\begin{equation}
\label{pdf_eq}
 p\left( w \right) = \left\{ {\begin{array}{*{20}c}
   {0, {\rm{\;\;\;when\;\;\;}}  \lambda_w\equiv 4\gamma (w-w_0)<0 \mbox{ else; }}  \\
{\sqrt{\frac{{2}}{{\pi}{\lambda_{w}}}}
{\cosh\left(\frac{\sqrt{\lambda_{w}}}{4\gamma^2}\right)}
\exp\left({-\frac{1+\lambda_{w}}{8\gamma^2}}\right),}
\end{array}} \right.
\end{equation}
where $w_{0} = - (4\gamma)^{-1} - \gamma$. The value and the sign of
parameter $\gamma$ controls the shape and the range of non-zero
values of $p(w)$. For $\gamma = 0$, the $p(w)$ reduces to that of a
centered Gaussian random process. For $\gamma>0$ ($\gamma<0$) a cut
off value $w_0$ exists and $p(w)$ gets asymmetric presenting long
tails in the positive (negative) axis (cf. Figure 3). The minimum
(in absolute sense) value of the cut-off is equal to $w_{0m} = \pm
1$ and corresponds to $\gamma = \mp 1/2$. Note, that the same
analytic expression for $p(w)$ can also be derived by noticing that
the non-Gaussian process $W$ can be re-written in terms of a scaled
and shifted non-central chi-squared random process with one degree
of freedom,
\begin{equation}
W = w_0+ \gamma \left( {\frac{Z}{{\sigma _0 }} + \frac{1}{{2\gamma
}}} \right)^2
\end{equation}
and using the expression of non-central chi-squared PDF along with
standard transform techniques of random variables.

In Fig.~\ref{fig:pdfplus}, we have plotted $p(w)$ choosing positive
values of $\gamma$ that range up to $\gamma=0.2$. The resulting
distributions exhibit the same behavior with the measured
distributions of plasma edge density fluctuations, as have been
observed in several magnetic confinement devices and presented here
in Fig.~\ref{fig:exppdf}.
Note, that the characteristic clustering of the PDF curves around
$0.05$ for values around $2.5$ which is experimentally observed (cf.
Fig. 1) is recovered by the distribution $p(w)$ (cf. Figs. 3 and 4)
- independently on the value of $\gamma>0$ - a characteristic that
is not recovered in the plots of PDF in Ref. \cite{Krommes}.
Similar features appear also in the PDF of the x--ray observations
associated with anomalous transport in accretion disks (cf. Fig. 8
of Ref. \cite{greenhogh}).

\begin{figure}
\begin{center}
\includegraphics[width=80mm]{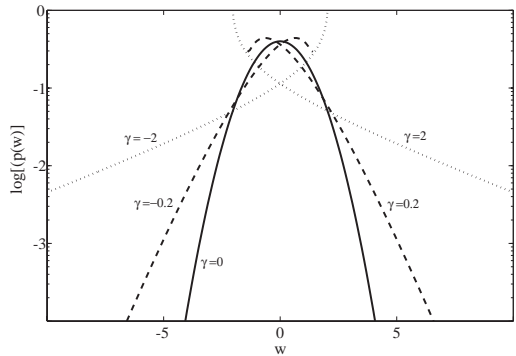} \vspace{-0.5cm}
\caption{{\small{The family of distributions $p(w)$, for different
values of $\gamma=-2,~-0.02,~0,~0.02\mbox{ and }2$. The symmetry
with respect to $\gamma$ is evident. As $\gamma$ increases the
absolute value of the cut--off decreases (increases) for
$\gamma<0.5$ ($\gamma>0.5$). }}} \label{fig:pdfall}
\end{center}
\end{figure}

\begin{figure}
\begin{center}
\includegraphics[width=80mm]{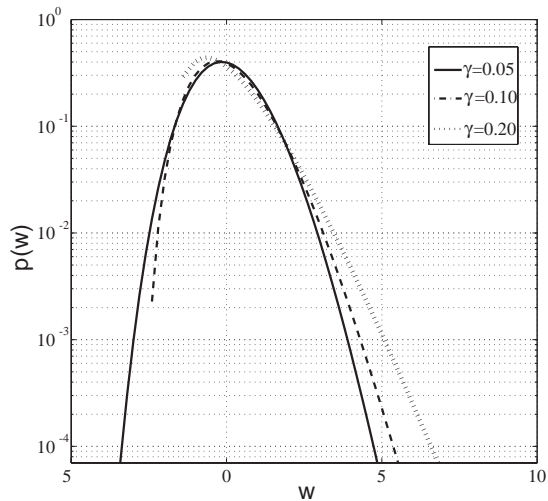} \vspace{-0.5cm}
\caption{{\small{The family of distributions $p(w)$, for
$\gamma=0.05,~0.1\mbox{ and }0.2$. The resulting PDFs recover the
universally observed distributions presented at Fig.
\ref{fig:exppdf}.  }}} \label{fig:pdfplus}
\end{center}
\end{figure}

Unlike, to the BHP distribution [13] which does not have any free
parameter, $p(w)$ depends on $\gamma$ such that its higher order
moments can receive multiple values. Furthermore, the existence of a
single free parameters allows $p(w)$ to be used for the fitting of
experimentally observed distributions.
The high order moments of multivariate Gaussian processes can be
determined by a simple method based on the Wicks theorem. All odd
moments are zero and all even moments can be reduced to homogeneous
polynomials. For the considered process $W$, the values of skewness
and kurtosis depend on the value of $\gamma$ and are given by
\cite{Lenschow},
\begin{equation} \label{sk_ku_eq}
S=2\gamma\frac{3+4\gamma^2}{(1+2\gamma^2)^{3/2}}\mbox{ and }
K=48\gamma^4\frac{1+\gamma^2}{(1+2\gamma^2)^2}
\end{equation}
For $\gamma=0$, the process $W$ is Gaussian and $K=S=0$. As
$\left|\gamma\right|$ increases kurtosis and skewness converge to
their extreme values $K_{max}=12$ and $S_{max}=\pm2^{3/2}$,
respectively. Noticeably, these values agree with the range of the
values reported in the Letter by Garcia et al \cite{Garcia}, in
which the authors investigate intermittent transport in plasma edge
using extensive numerical simulations that take into account the
presence of SOL.

Using the parabolic ansatz, $K=aS^2+b$ for the parametric relations
in Eq. (\ref{sk_ku_eq}), it can be easily found that
\begin{equation} \label{sk_ku_eq2}
a(\gamma)=12\frac{(1+\gamma^2)(1+2\gamma^2)}{(3+4\gamma^2)^2},
\end{equation}
and $b=0$. The function $a(\gamma)$ converges rapidly to the value
$a=3/2$, which is exactly the value of the universally observed
parabolic scaling, reported in Refs. \cite{Labit,Krommes,Sura}.

It is interesting to compare the derived $K-S$ curve of Eq.
(\ref{sk_ku_eq}), to that attributed to a quadratic product of two
central Gaussian processes $Z_1,~Z_2$, i.e.
$W_{\Gamma}=\gamma_{\Gamma}Z_1Z_2$. Here $\gamma_{\Gamma}$ denotes
the correlation between the Gaussian fields. For $Z_1=Z_2$ and
$\gamma_{\Gamma}=1$, the PDF of $W_{\Gamma}$ is simply the
chi-squared distribution with one degree of freedom. The associated
PDF $p(W_{\Gamma})$ has been presented in Ref. \cite{Carreras}, and
the corresponding values of skewness and kurtosis were found equal
to:
\begin{equation} \label{sk_ku_eq_Gamma}
S_{\Gamma}=-2\gamma_{\Gamma}\frac{3+\gamma_{\Gamma}^2}{(1+\gamma_{\Gamma}^2)^{3/2}}
\mbox{ and }
K_{\Gamma}=6\frac{1+6\gamma_{\Gamma}^2+\gamma_{\Gamma}^4}{(1+\gamma_{\Gamma}^2)^2},
\end{equation}
respectively. The parabolic relation, $K=a_{\Gamma}S^2+b_{\Gamma}$,
between $S_{\Gamma}$ and $K_{\Gamma}$, results to the derivation of
the following coefficients:
\begin{equation} \label{Gamma_sc}
a_{\Gamma}=\frac{6(1+\gamma_{\Gamma}^2)}{(3+\gamma_{\Gamma}^2)^2}\mbox{
and } b_{\Gamma}=6.
\end{equation}
Note that for $\left|\gamma_{\Gamma}\right|=1$, the values of
kurtosis and skewness are equal to $K_{max}$ and $S_{max}$, while
for $\gamma_{\Gamma}=0$, are equal to $6$ and $0$, respectively.

\begin{figure}
\begin{center}
\includegraphics[width=80mm]{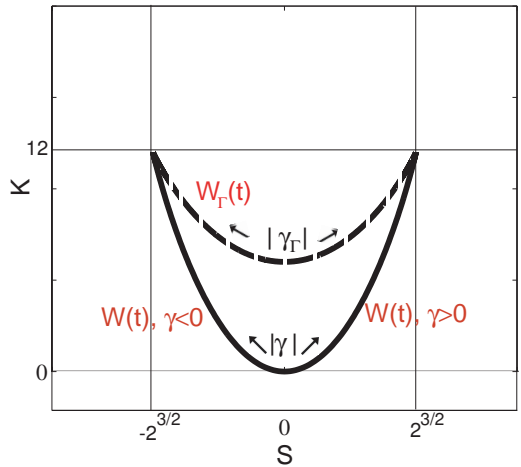}
\caption{{\small{The $K-S$ curve for stochastic processes described
by quadratic polynomial of gaussians. The solid line corresponds to
$W(t)$ and the dashed line to $w_{\Gamma}(t)$ process.}}}
\label{fig:KS_curve}
\end{center}
\end{figure}

Equations (\ref{sk_ku_eq}) and (\ref{Gamma_sc}) define a closed
curve in the $K-S$ space for continuous values of the parameters
$\gamma$ and $\gamma_{\Gamma}$ respectively (see Fig.
\ref{fig:KS_curve}). The low boundary of the curve corresponds to
extreme bursty processes $W$ given by Eq. (1) and distributed
according to $p(w)$, while the upper boundary corresponds to
$W_{\Gamma}$ which is distributed according to the ``local flux PDF"
(see Eq. (7) in Ref. \cite{Carreras}). The clustering of the $K-S$
values around the $K=1.5S^2$ curve is patently seen in Fig. 1 of
Reference \cite{Labit} and in Fig. 3 of Reference \cite{Sura}.

It is straightforward to show numerically, that for non--Gaussian
processes (described by second order polynomials of Gaussians) the
``cloud" of $K-S$ points fall into (not shown here) the closed
curve. The latter follow a parabolic trend -- similar to the
experiments \cite{sattin} -- with scaling that depends on the
selection of the polynomial coefficients. Furthermore, for processes
described by higher order Gaussian polynomials the ``cloud" follows
also parabolic trend including much higher values of $K$ and $S$.

In conclusion, a univariate stochastic model for the description of
extreme bursty fluctuations $W$ based on generic properties of
quadratic non-linearities is presented. The model and the associated
probability distribution function describe statistical properties of
the SOL density fluctuations which are governed by the $E \times B$
convection. The universal character of the stochastic process stems
from the fact that the associated extreme PDF recovers properties of
variabilities (density/sea temperature/x-ray intensity) observed at
boundary regions (SOL/sea surface/accretion edge) of different
physical processes that are characterized by convection (electric
drift/ocean current/rotation). The proposed univariate stochastic
model describes the statistical characteristics of these relaxation
phenomena at a state of extreme statistical behavior. It is evident
that the parabolic relation between $S$ and $K$ when observed
provide relevant information about the underlying processes.

IS acknowledges fruitful discussions with Yu. Khotyaintsev, F.
Lepreti and A. Anastasiadis. DCN acknowledges support from the Oak
Ridge National Laboratory, managed by UT-Battelle, LLC, for the U.S.
This work was supported under the Contract of Association ERB 5005
CT 99 0100 between the European Atomic Energy Community and the
Hellenic Republic. The content of the publication is the sole
responsibility of its author(s) and it does not necessarily
represent the views of the Commission or its services.

%

\end{document}